\documentclass[a4paper]{jpconf}
\usepackage{graphicx}

\newcommand{\beqn}{\begin{eqnarray}}
\newcommand{\eeqn}{\end{eqnarray}}

\def\beq{\begin{equation}}
\def\eeq{\end{equation}}
\def\beeq{\begin{eqnarray}}
\def\eeeq{\end{eqnarray}}

\begin{document}

\title{Exploring the polarization of gluons in the nucleon}

\author{Marco Stratmann}

\address{Radiation Laboratory, RIKEN, 2-1 Hirosawa, Wako, 
Saitama 351-0198, Japan}

\ead{marco@ribf.riken.jp}

\author{Werner Vogelsang\footnote{Invited plenary talk presented 
at the ``Second Meeting of the APS Topical Group on Hadronic Physics'',
Nashville, Tennessee, October 22-24, 2006}}

\address{BNL Nuclear Theory,
Brookhaven National Laboratory, Upton, NY 11973, USA}

\ead{wvogelsang@bnl.gov}

\begin{abstract}
We give an overview of the current status of investigations of the 
polarization of gluons in the nucleon. We describe some of the 
physics of the spin-dependent gluon parton distribution and 
its phenomenology in high-energy polarized hadronic scattering.
We also review the recent experimental results.\\[-8.6cm]
\hspace*{11.3cm}BNL-NT-07/12\\[7.2cm]
\end{abstract}

\section{Introduction}

\begin{figure}[t]
\includegraphics[width=15pc,angle=90]{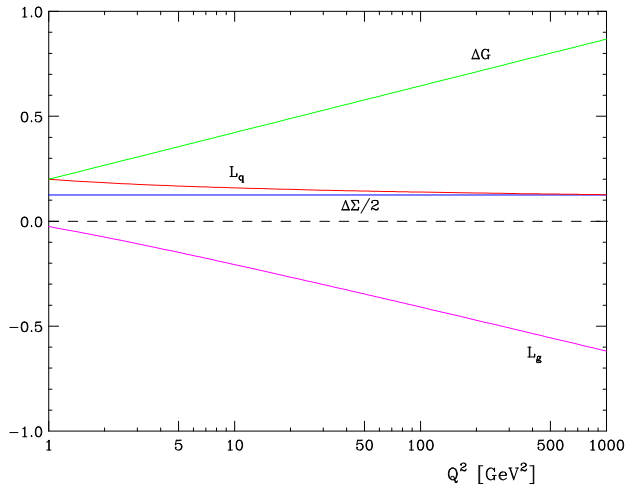}\hspace{2pc}%
\begin{minipage}[b]{17pc}\caption{LO toy calculation of the $Q^2$-evolution
of the contributions to the proton spin. \label{fig1}}
\end{minipage}
\end{figure}

For many years now, spin has played a very prominent
role in QCD. The field of QCD spin physics has been driven by
the hugely successful experimental program of polarized 
deeply-inelastic lepton-nucleon scattering (DIS)~\cite{hughes}. 
One of the most 
important results of this program has been the finding that the 
quark and anti-quark spins (summed over all flavors) provide only 
about a quarter of the nucleon's spin, $\Delta \Sigma\approx 0.25$ 
in the proton helicity sum rule~\cite{helsr,helsr1a,helsr1}
\begin{equation} 
\frac{1}{2}=\frac{1}{2}\Delta \Sigma(Q^2) + \Delta G(Q^2)+
L_q(Q^2)+L_g(Q^2) \; , \label{ssr}
\end{equation} 
implying that sizable contributions to the nucleon spin should come 
from the gluon spin contribution $\Delta G(Q^2)$, or from 
orbital angular momenta $L_{q,g}(Q^2)$ of partons. Here, $Q$
is the resolution scale at which one probes the nucleon. The
$Q^2$-dependence of the various contributions to 
the proton spin is predicted in perturbative QCD through its 
evolution equations~\cite{helsr1,dglap,kod}. To
lowest order (LO), the quark and anti-quark spin contribution 
$\Delta \Sigma(Q^2)/2$ does not depend on $Q^2$. Figure~\ref{fig1}
shows a LO toy calculation of the $Q^2$-evolution of the contributions 
to the proton spin in Eq.~(\ref{ssr}), assuming that at an initial 
scale $Q_0=1$~GeV we have $\Delta \Sigma =0.25$, $\Delta G = L_q=0.2$, 
$L_g= -0.025$. The rise of $\Delta G\propto\log (Q^2)$ or
$1/\alpha_s(Q^2)$ (compensated by an opposite evolution of $L_g$)
is an important prediction of QCD and awaits experimental testing.
For the initial conditions chosen here, the evolution leads to large
positive values of $\Delta G$. We note that at asymptotic $Q^2$ 
the total quark and gluon angular momenta, $\frac{1}{2}\Delta \Sigma+
L_q$ and $\Delta G + L_g$, respectively, become roughly 
equal~\cite{helsr1}. 

To determine the gluon spin contribution on the right-hand-side of 
Eq.~(\ref{ssr}) has become a major focus of the field. Like $\Delta \Sigma$,
it can be probed in polarized high-energy scattering. Several current 
experiments are dedicated to a direct determination of the spin-dependent 
gluon distribution $\Delta g(x,Q^2)$,
\begin{equation} \label{qdef}
\Delta g(x,Q^2)  \equiv g^+(x,Q^2)-g^-(x,Q^2)   \; ,
\end{equation} 
where $g^+$ ($g^-$) denotes the number density of gluons in a longitudinally
polarized proton with same (opposite) sign of helicity as the proton's,
and where $x$ is the gluon's light-cone momentum fraction.
The field-theoretic definition of $\Delta g$ is 
\begin{equation}
\Delta g (x,Q^2) =
\frac{i}{4\pi\,x\, P^+} \int d\lambda \, {\rm e}^{i\lambda x P^+}\,
\langle P,S| G^{+\nu}(0)\, \tilde{G}^+_{\;\;\nu} 
(\lambda n)|P,S\rangle\Big|_{Q^2} \;, \label{opme}
\end{equation}
written in $A^+=0$ gauge. $G^{\mu\nu}$ is the QCD field strength tensor, 
and $\tilde{G}^{\mu\nu}$ its dual. The integral of $\Delta g(x,Q^2)$ 
over all momentum fractions $x$ becomes a local operator only in $A^+=0$ 
gauge and then coincides with $\Delta G(Q^2)$~\cite{helsr,helsr2}. 
The COMPASS experiment at CERN and the HERMES 
experiment at DESY attempt to access $\Delta g(x,Q^2)$ 
in charm- or high-$p_T$ hadron final states in photon-gluon fusion 
$\gamma^{\ast}g\to q\bar{q}$. A new milestone has been reached with the 
advent of the first polarized proton-proton collider, RHIC at 
BNL~\cite{rhicrev,rhicrev1}. RHIC will provide precise and detailed
information on $\Delta g$, over a wide range of $x$ and $Q^2$, and
from a variety of probes.

\section{Model estimates of $\Delta g$}

Before we discuss in some detail the phenomenology of $\Delta g$ in
polarized high-energy scattering, let us briefly address some of the 
available theoretical expectations for $\Delta g$ and its integral. As was 
first pointed out in~\cite{jaffeDg}, it is possible to estimate
the operator matrix element corresponding to $\Delta G$ 
in non-relativistic quark and bag models. In such models, 
for example, baryon mass splittings result from lowest-order 
exchange of transverse gluons, and the associated forces are 
spin-dependent. One obtains estimates~\cite{Barone,JiChen} 
for $\Delta G(Q^2\approx 1\;
{\mathrm{GeV}}^2)$ of about $0.2$ to $0.3$. In a sense, these are ``natural''
values since they are of the order of the proton spin itself. 
Very recently, for the first time model calculations of the 
$x$-dependence of $\Delta g$ have been presented~\cite{JiChen}. The 
resulting distribution is positive everywhere and of moderate size.
The more and more precise experimental constraints on $\Delta g$ will 
likely motivate further model investigations, which ultimately might
lead to new insights into QCD. Likewise, it is to be hoped that
lattice calculations, which are becoming ever more powerful,
will be able to address gluonic observables in nucleon 
structure in the future~\cite{Negele}.

Other considerations, based in part on perturbation theory, 
led to the prediction of a very large gluon polarization in the 
nucleon. The peculiar evolution pattern of $\Delta G(Q^2)\propto 
1/\alpha_s(Q^2)$ visible in Fig.~\ref{fig1} inspired ideas~\cite{as}
that a reason for the experimentally found small size of the proton's
axial charge should be sought in a ``shielding'' of the quark spins due 
to a particular perturbative part of the DIS process $\gamma^{\ast}g
\to q\bar{q}$. The associated cross section is of order $\alpha_s(Q^2)$,
but the $Q^2$-evolution of $\Delta G(Q^2)$ would 
compensate this suppression. We note that this interpretation of
the axial charge, however, corresponds to a particular choice of 
factorization scheme. To be of any phenomenological relevance, 
such ``anomalous'' models would require a very large positive 
gluon spin contribution, $\Delta G>1.5$, even at a low scale of 1~GeV or 
so. As we shall see below, initial experimental data now appear to make 
such a scenario very unlikely.

\section{$\Delta g$ and scaling violations in polarized DIS}

\begin{figure}[h]
\includegraphics[width=18pc]{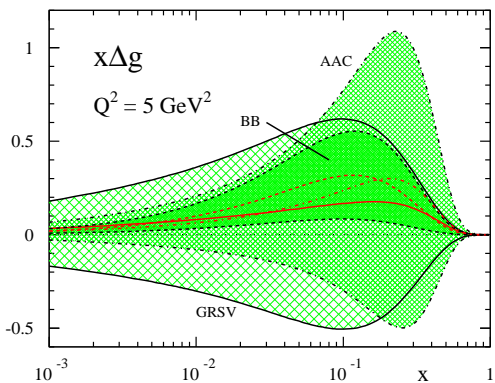}\hspace{2pc}%
\begin{minipage}[b]{18pc}
\caption{Results for $x\Delta g(x,Q^2=5~{\rm GeV}^2)$ 
from several analyses~\cite{bb,grsv,aac} 
of polarized DIS. The various bands indicate ranges in $\Delta g$ that 
were deemed consistent with the DIS scaling violations 
in these analyses. From~\cite{rhicrev1}. \label{gluondist}}
\end{minipage}
\end{figure}

In principle, a clean determination of $\Delta g(x,Q^2)$ is 
possible by investigating scaling violations of the spin-dependent
proton structure function $g_1(x,Q^2)$ which is measured in polarized
DIS. To leading order of QCD, $g_1$ can be written as 
\begin{equation} \label{g1lo}
g_1 (x,Q^2) = \frac{1}{2} \sum_q e_q^2 \left[ \Delta q(x,Q^2) + \Delta 
\bar{q}(x,Q^2) \right] \; , 
\end{equation}
where the $\Delta q$ and $\Delta \bar{q}$ are the quark and anti-quark
helicity distributions. QCD predicts the
$Q^2$-dependence of the densities through the spin-dependent
DGLAP evolution equations~\cite{dglap}:
\begin{equation} \label{dglapeq}
\frac{d}{d \ln Q^2} \left( \begin{array}{c}
\! \Delta q \!
\\ \! \Delta g \! \end{array} \right)(x,Q^2) 
= \int_x^1\frac{dz}{z}\,\left( \begin{array}{cc}
\! \Delta P_{qq}(\alpha_s(Q^2),z) &  \Delta  P_{qg}(\alpha_s(Q^2),z) \! \\
\! \Delta P_{gq}(\alpha_s(Q^2),z) & \Delta P_{gg}(\alpha_s(Q^2),z) \!
\end{array} \right) \;
\left( \begin{array}{c}
\!  \Delta q \! \\ \!  \Delta g \!
\end{array} \right) \left( \frac{x}{z},Q^2 \right) \; ,
\end{equation}
the $\Delta P_{ij}$ 
are the spin-dependent ``splitting functions''~\cite{dglap,mvn}
which are evaluated in QCD perturbation theory. As one can see,
$\Delta g$ contributes to the scaling violations of $g_1$.
Nonetheless, $\Delta g$ has been left virtually 
unconstrained (see, for example,~\cite{bb,grsv,aac,ddf,lss}) 
by the scaling violations observed
experimentally in polarized DIS. This is due to the very limited lever 
arm in $Q^2$ of the fixed-target experiments. Figure~\ref{gluondist}
shows current theoretical ``uncertainty bands'' for $\Delta g$ from
DIS scaling violations. At best, a tendency toward a positive
$\Delta g$ is seen. We note that a recent new analysis by the
COMPASS collaboration~\cite{mallot} using their latest deuteron
DIS data~\cite{compassg1d} finds two 
``allowed'' regions for $\Delta g$, one with positive, one with
negative gluon polarization. 
Clean and precise extractions of $\Delta g(x,Q^2)$ over
a wide range of $x$ and $Q^2$ from scaling violations of $g_1$ would 
become possible at a polarized electron-ion collider, EIC~\cite{erhicrev},
thanks to its vastly larger kinematic reach.

\section{Access to $\Delta g$ in polarized proton-proton 
scattering at RHIC}

The measurement of gluon polarization 
in the proton is a major focus and strength of RHIC~\cite{rhicrev,rhicrev1}. 
The basic concept that underlies most of spin physics at RHIC
is the factorization theorem~\cite{fact}. It states that large
momentum-transfer reactions may be factorized into
long-distance pieces that contain the desired information on the
spin structure of the nucleon in terms of its {\em universal} parton
densities, and parts that are
short-distance and describe the hard interactions of the
partons. The latter can be evaluated using perturbative QCD.
As an example, we consider the double-spin asymmetry for 
the reaction $pp\to \pi X$, 
\begin{equation}
A_{\mathrm{LL}}\equiv \frac{\sigma^{++}-\sigma^{+-}}{\sigma^{++}+\sigma^{+-}}
\equiv \frac{\Delta \sigma}{\sigma} \; ,
\end{equation}
where the superscripts denote helicities of the initial protons.
We assume the pion to be produced at high transverse 
momentum $p_T$, ensuring large momentum transfer. Then, up to 
corrections suppressed by inverse powers of $p_T$:
\begin{equation}
\label{eq:eq2}
d\Delta \sigma =\sum_{abc}\, 
\Delta f_a \,\otimes \,\Delta f_b  \,\otimes\,
d\Delta \hat{\sigma}_{ab}^{c} \,\otimes \,D_c^{\pi} 
\end{equation}
for the polarized cross section, where $\otimes$ denotes a convolution. 
The $\Delta f_i$ are the polarized parton distributions, and $D_c^{\pi}$ 
the pion fragmentation functions. The sum in Eq.~(\ref{eq:eq2}) 
is over all  contributing partonic channels $a+b\to c + X$, with
$d\Delta \hat{\sigma}_{ab}^{c}$ the associated spin-dependent partonic cross
section. In general, a leading-order estimate of (\ref{eq:eq2})
merely captures the main features, but does not usually provide a 
quantitative understanding. 
Only with knowledge of the next-to-leading order (NLO)
QCD corrections to the $d\Delta \hat{\sigma}_{ab}^{c}$
can one reliably extract information on the parton distribution functions 
from the reaction. 

Several different processes will be investigated at 
RHIC~\cite{rhicrev,rhicrev1} that are very sensitive to 
gluon polarization: high-$p_T$ prompt photons 
$pp\to \gamma  X $, jet or hadron production $pp\to {\rm jet}\,X$, 
$pp\to h X$, and heavy-flavor production $pp\to (Q\bar{Q}) X$.
An important role for the determination of $\Delta g$ will be played
by measurements of  two-particle, jet-jet (or hadron-hadron) and 
photon-jet correlations. For these, at the leading order
approximation, the hard-scattering subprocess kinematics can be 
calculated directly on an event-by-event basis, giving an estimate 
of the gluon momentum fraction~\cite{les}. 
In addition, besides the current $\sqrt{s}=200$~GeV, also 
$\sqrt{s}=500$~GeV will be available at RHIC at a later stage. All
this will allow to determine $\Delta g(x,Q^2)$ in various
regions of $x$, and at different scales. Essentially all tools 
are in place now for treating the spin-dependent reactions relevant 
at RHIC at NLO~\cite{jssv,jsvj,jsvj1,jsvj2}. 
\begin{figure}[p]
%\vspace*{-1.2cm}
\hspace*{-0.3cm}
\includegraphics[width=18pc]{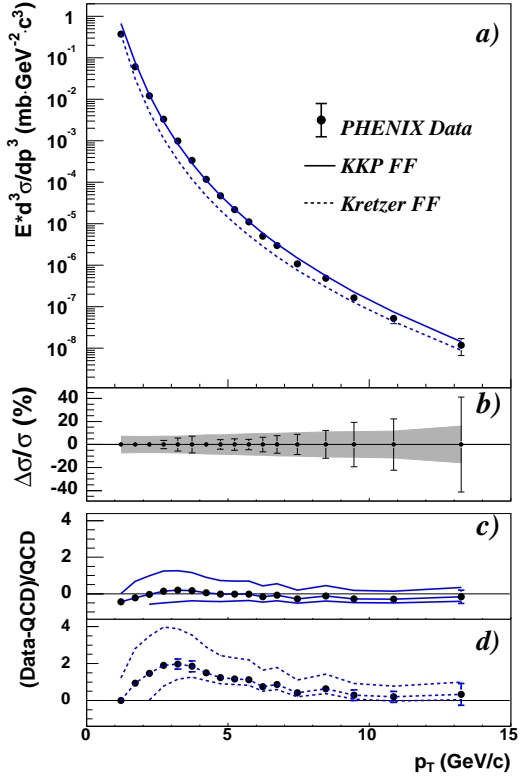}
\hspace{0.8cm}
\includegraphics[width=17pc]{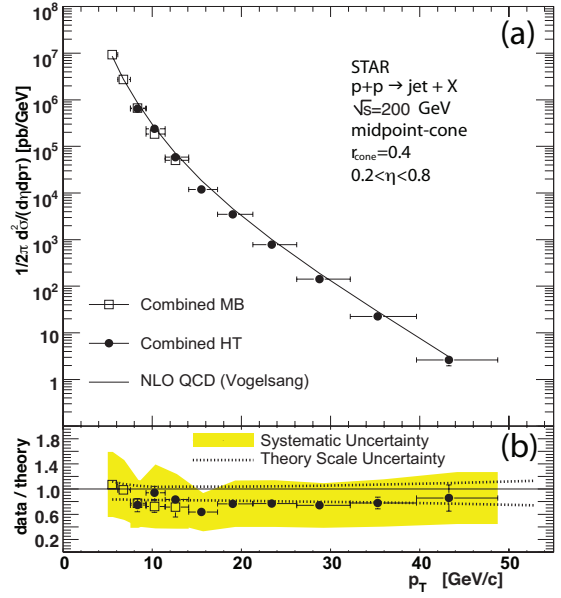}

\vspace*{0.6cm}
\hspace{0.0cm}
\includegraphics[width=18pc]{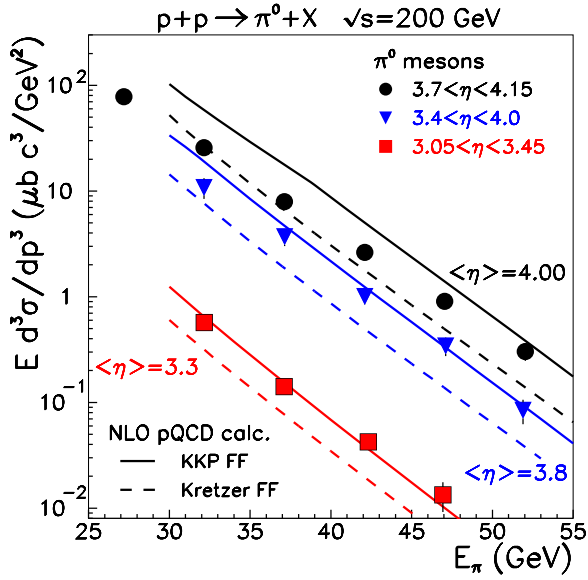}
\hspace{1.0cm}
\vspace*{3mm}
\includegraphics[width=18pc]{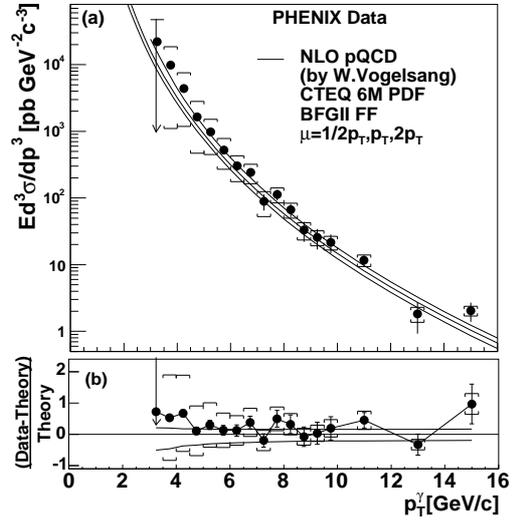}
\caption{Data for the cross section for single-inclusive 
$\pi^0$ production $pp\to \pi^0 X$ at $\sqrt{s}=200$~GeV
at mid-rapidity from PHENIX (upper left,~\cite{cross_phenix}) 
and at forward rapidities from STAR (lower left, \cite{cross_star}),
for mid-rapidity jet production from STAR (upper 
right, \cite{star_jets}), and for mid-rapidity prompt-photon
production from PHENIX (lower right, \cite{phenix_photons}).
The lines show the results of the corresponding next-to-leading order 
calculations~\cite{jssv,jsvj1,jsvj2}.
\label{fig:rhicdata}}
\end{figure}

\begin{figure}[h]
\includegraphics[width=18pc]{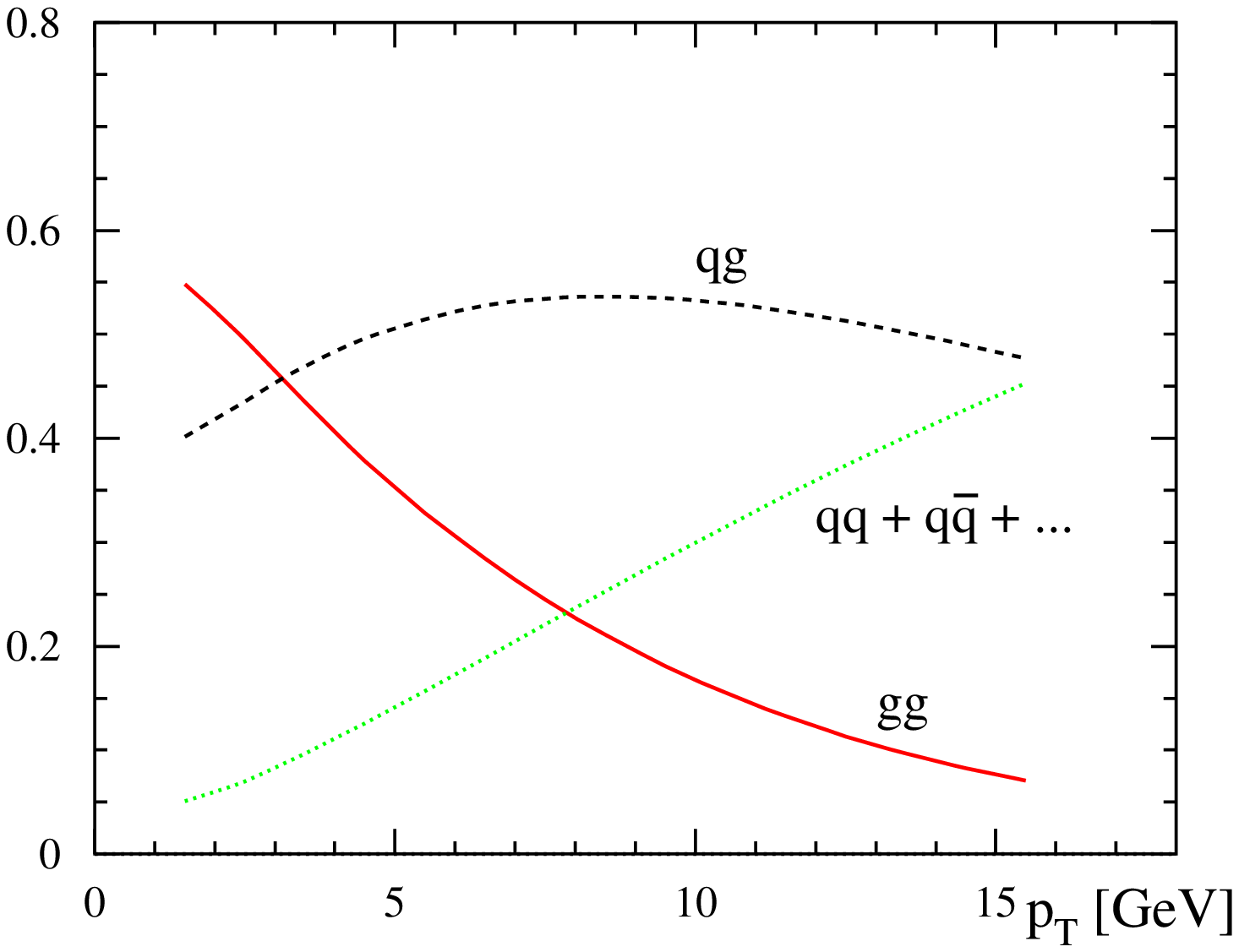}\hspace{2pc}%
\begin{minipage}[b]{18pc}
\caption{Relative contributions to the mid-rapidity NLO cross section 
for $pp\to \pi^0 X$ at $\sqrt{s}=200$~GeV from $gg$, $qg$, and $qq$ initial 
states~\cite{rhicrev1}.  \label{rhicdata1}}
\end{minipage}
\end{figure}

We emphasize that there have already been results from RHIC that demonstrate
that the NLO framework is very successful. Figure~\ref{fig:rhicdata} shows 
comparisons of data from PHENIX and STAR for single-inclusive
cross sections for $\pi^0$~\cite{cross_phenix,cross_star}, 
jets~\cite{star_jets} and photons~\cite{phenix_photons} with corresponding 
NLO calculations~\cite{jssv,jsvj1,jsvj2,other}. As can be seen, 
the agreement is overall 
excellent. We note that an agreement between data and NLO calculations
like the one seen in Fig.~\ref{fig:rhicdata} is not found in the 
fixed-target regime~\cite{bs} (it has recently been shown that in this 
regime large logarithmic terms at yet higher orders are important and 
need to be resummed for a more successful theoretical 
description~\cite{ddfwv}). In Fig.~\ref{rhicdata1} we decompose 
the NLO mid-rapidity $\pi^0$ cross section into the relative contributions 
from the various two-parton initial states~\cite{rhicrev1}. It is evident that 
processes with initial gluons dominate. 

\begin{figure}[h]
%\vspace*{-0.1cm}

\includegraphics[width=18pc]{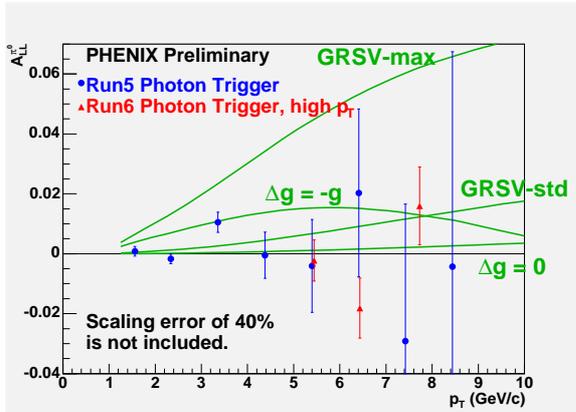}
\hspace{0.8cm}
\includegraphics[width=19pc]{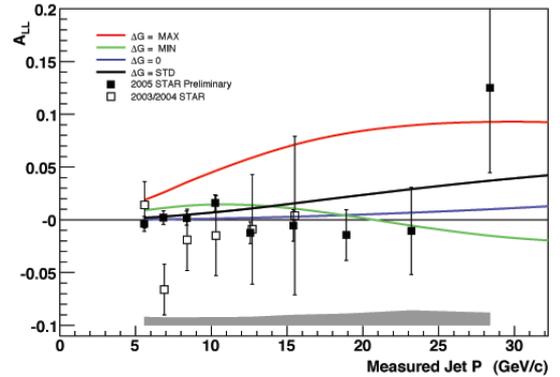}

\caption{Data for the double-spin asymmetry for mid-rapidity 
single-inclusive $\pi^0$ production at $\sqrt{s}=200$~GeV
from PHENIX~\cite{allpionphenix} (left), and for jet production
from STAR~\cite{star_jets} (right), compared to NLO
predictions for several polarized gluon distributions of~\cite{grsv}. 
\label{fig:rhicalldata}}
\end{figure}

The results shown in Fig.~\ref{fig:rhicdata}
give confidence that the theoretical NLO framework may be 
used to determine the spin-dependent gluon density from 
RHIC data. Results for $A_{\mathrm{LL}}$ in $pp\to \pi X$ are now available
from PHENIX~\cite{allpionphenix}, and $A_{\mathrm{LL}}$ for single-inclusive
jet production has been measured by STAR~\cite{star_jets}. 
The results are shown in Fig.~\ref{fig:rhicalldata}.
The curves shown in Fig.~\ref{fig:rhicalldata} 
represent the $A_{\mathrm{LL}}$ values calculated at NLO for a range of 
gluon distributions from~\cite{grsv}, from a suggested very large 
positive gluon polarization (``GRSV-max'')
with an integral $\Delta G=1.9$ at scale $Q=1$~GeV, 
to a "maximally" negative gluon polarization, (``$\Delta g=-g$''),
for which $\Delta G(1\,\mathrm{GeV}^2)= -1.8$. These
two distributions span the ``GRSV-band'' shown in Fig.~\ref{gluondist}.
The curves labeled  ``GRSV-std'' represent the best fit of~\cite{grsv} to 
the polarized DIS data (solid line in Fig.~\ref{gluondist}), which has 
a more ``natural'' $\Delta G(1\, \mathrm{GeV}^2)$ of about 0.4, and
the results for ``$\Delta g=0$'' correspond to 
very little gluon polarization, $\Delta G(1\,\mathrm{GeV}^2)= 0.1$.
One can see that the data are already discriminating between the various
$\Delta g$ distributions. A very large gluon distribution, as proposed 
in the context of the ``anomaly scenario'' (see discussion above) and 
corresponding roughly to the curves labeled ``GRSV max'', appears to be 
strongly disfavored.

\begin{figure}[h]
\includegraphics[width=18pc]{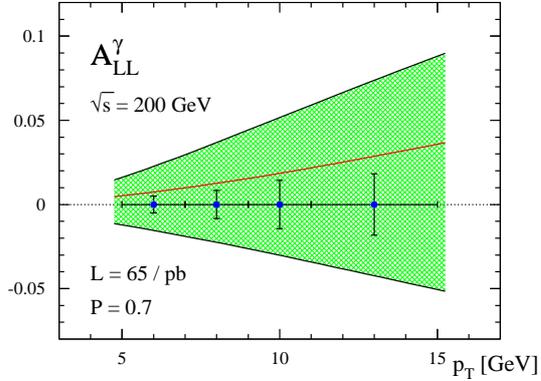}\hspace{2pc}%
\begin{minipage}[b]{18pc}\caption{``NLO theory band'' for 
mid-rapidity prompt-photon production 
at PHENIX. The band illustrates the current ``uncertainty'' due 
to $\Delta g$, using the GRSV $\Delta g$ band shown in Fig.~\ref{gluondist}.
The ``errors'' are projections. Taken from~\cite{rhicrev1}. 
\label{fig:gluon1c}}
\end{minipage}
\end{figure}

Figure~\ref{fig:gluon1c} shows NLO predictions~\cite{jsvj1}  
for the double-longitudinal 
spin asymmetry $A_{\mathrm{LL}}$ for the reaction $pp\to\gamma X$ at RHIC, 
based on the ``gluon uncertainty'' band displayed in Fig.~\ref{gluondist}. 
Prompt photons are much less copiously produced than pions at RHIC,
resulting in larger statistical uncertainties. The measurement of this
asymmetry will therefore take some time at RHIC. Nonetheless, the
reaction $pp\to\gamma X$ is of great importance because of its 
direct sensitivity to $\Delta g$ through the clean ``Compton-like'' process
$qg\to \gamma q$. The spin asymmetry for this reaction is linear in 
$\Delta g$ and therefore directly determines the sign of the distribution.
The plot also shows the experimental uncertainties expected at RHIC 
(PHENIX) for 65/pb collected luminosity~\cite{rhicrev1}. 

\section{$\Delta g$ from photon-gluon fusion}

\begin{figure}[h]
\includegraphics[width=18pc]{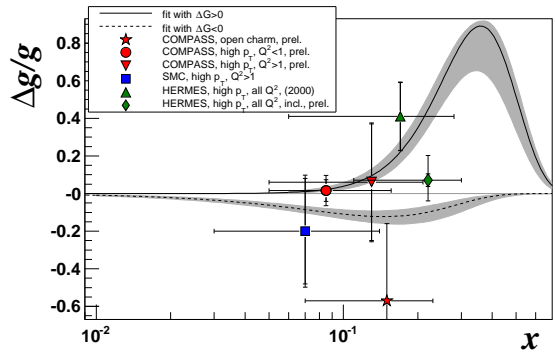}\hspace{2pc}%
\begin{minipage}[b]{18pc}\caption{SMC~\cite{smcdg}, 
COMPASS~\cite{compass}, and HERMES~\cite{hermesdg}
results for gluon polarization, extracted from photon-gluon fusion.
Taken from~\cite{mallot}. \label{fig:gluon1d}}
\end{minipage}
\end{figure}

A way to access $\Delta g$ in lepton-nucleon scattering 
is to measure final states that select the photon-gluon fusion process.
These are heavy-flavor production, $\ell p\to c\bar{c} X $,
and single- or di-hadron production, $\ell p\to h X$ or 
$\ell p\to h_1 h_2 X $, where the hadrons have large transverse momentum. 
Figure~\ref{fig:gluon1d} compiles the current 
results~\cite{mallot,smcdg,compass,hermesdg} for extractions of
$\Delta g$ from these reactions. We note that, unlike at RHIC, the 
success of the perturbative-QCD hard-scattering description has not been 
established for these observables in the kinematic regimes of interest here. 
Also, the translation of the measured spin asymmetry into $\Delta g$ 
at a certain single momentum fraction, currently only possible at
leading order, is fraught with large uncertainties.

\section{Global Analysis}
\begin{figure}[t]
\includegraphics[width=32pc]{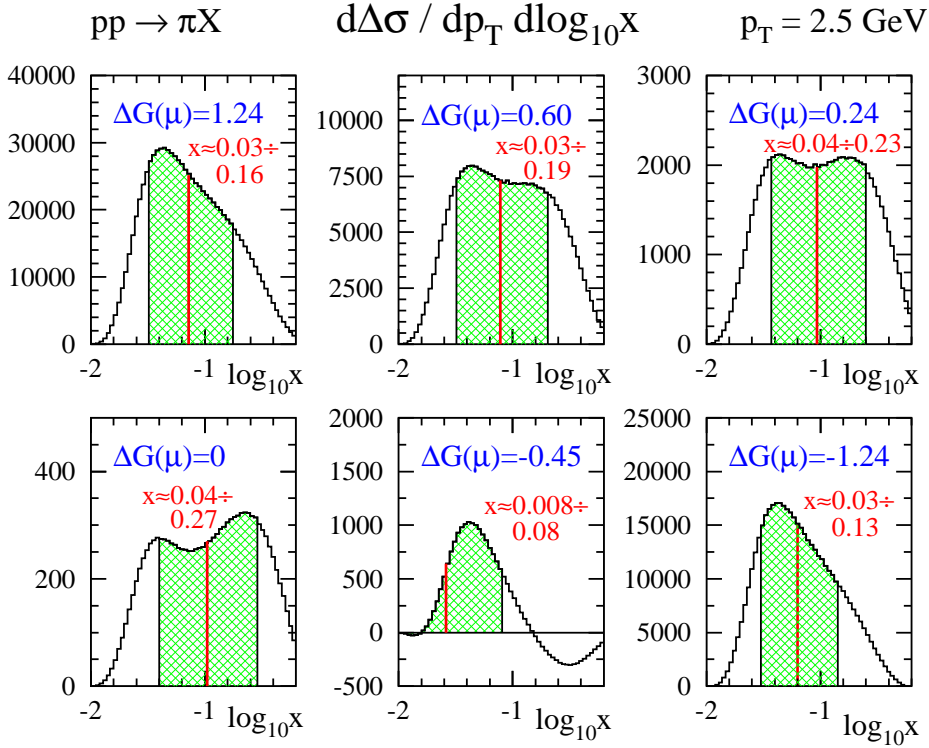}\hspace{2pc}%
\caption{NLO $d\Delta\sigma/dp_Td\log_{10}(x)$ (arbitrary
normalization) for the reaction
$pp\to \pi^0 X$ at RHIC, for $p_T=2.5\,\mathrm{GeV}$ and six 
different values for $\Delta G (\mu^2)$ at $\mu\approx 0.4$~GeV~\cite{grsv}. 
The shaded areas denote in each case the $x$-range dominantly 
contributing to $d\Delta\sigma$. From~\cite{DIS06}. \label{xdistr}}
\end{figure}

The eventual determination of gluon polarization will require 
consideration of all existing data through a ``global analysis'' that
makes simultaneous use of results for all probes, from RHIC and from
lepton scattering. The technique is to optimize the agreement between 
measured spin asymmetries, relative to the accuracy of the data, and the 
theoretical spin asymmetries, by minimizing the associated $\chi^2$ function
through variation of the shapes of the polarized parton distributions. 
The advantages of such a full-fledged global analysis program are manifold:
(1) The information from the various reaction channels
is all combined into a single result for $\Delta g(x)$. (2)
The global analysis effectively deconvolutes the experimental
information, which in its raw form is smeared over the fractional gluon 
momentum $x$, and fixes the gluon distribution at definite values of $x$.
Figure~\ref{xdistr} highlights the importance of this. The figure
shows~\cite{DIS06} 
the contributions of the various regions in gluon momentum fraction
to the mid-rapidity spin-dependent cross section for $pp\to \pi^0 X$ at 
RHIC, for six different sets of polarized parton distributions~\cite{grsv}
mostly differing in the gluon distribution. The pion's transverse
momentum was chosen to be $2.5$~GeV. One can see that the 
distributions are very broad, and that the $x$-region that is
mostly probed depends itself on the size and form of the polarized
gluon distribution. This makes it very difficult to assign a 
good estimate of the gluon momentum fraction to a data point at
a given pion transverse momentum. The global analysis solves this
problem.

The further advantages of a global analysis are:
(3) State-of-the-art (NLO) theoretical calculations can be used without 
approximations. (4) It provides a framework to determine an error on the 
gluon polarization. (5) Correlations with other experiments, to be 
included in $\chi^2$ and sensitive to degrees of freedom different 
from $\Delta g$, are automatically respected.
Global analyses of this type have been developed very successfully 
over many years for unpolarized parton densities.
Examples of early work on global analyses of RHIC-Spin and polarized 
DIS data in terms of polarized parton distributions are~\cite{aac1,df1,sv}. 

\section{Conclusions and Outlook}
\begin{figure}[h]
\includegraphics[width=18pc]{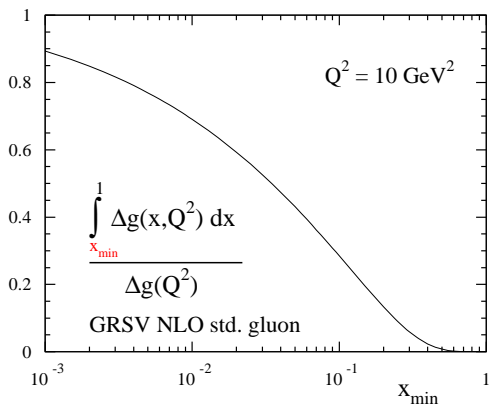}\hspace{2pc}%
\begin{minipage}[b]{18pc}\caption{``Running integral''
$\int_{x_{\mathrm{min}}}^1 dx \Delta g(x,Q^2=10\;{\mathrm{GeV}}^2)$,
normalized to the full integral, for the gluon distribution in the NLO
GRSV ``standard'' set~\cite{grsv}, as a function of $x_{\mathrm{min}}$.
\label{fig:gluon1f}}
\end{minipage}
\end{figure}

While the initial data from RHIC and from the dedicated studies
in lepton scattering shown above point to a small or moderate size 
of the gluon polarization in the $x$-region currently accessible, 
statements about the gluon contribution to the 
proton spin, $\Delta G$, are really not possible yet and will require 
the global analysis
just described. A crucial issue will eventually be the behavior of
the extracted $\Delta g(x)$ at the smallest accessible $x$, which are 
reachable in 500~GeV running at RHIC and in correlation studies involving 
final states produced at forward angles~\cite{les}. It is possible that a 
significant contribution to $\Delta G$ comes from relatively small $x$.
As one example, we show in Fig.~\ref{fig:gluon1f} the ``running integral''
$\int_{x_{\mathrm{min}}}^1 dx \Delta g(x,Q^2)$ at $Q^2=10$~GeV$^2$,
normalized to the full integral $\Delta G(Q^2)$, 
for the gluon distribution in the NLO
GRSV ``standard'' set~\cite{grsv}. As one can see, at this scale
about 30\% of the integral come from $x\leq 10^{-2}$. RHIC will
likely be able to constrain $\Delta g$ down to values somewhat
smaller than that, but the example shows that it might eventually
be necessary to push to $x\leq 10^{-3}$ and below. This
could be achieved at a high-energy polarized electron-proton 
collider~\cite{erhicrev}.

\ack
We are grateful to Barbara J\"{a}ger
for collaboration on some of the topics presented here. 
W.V.~thanks the U.S. Department of Energy (contract number 
DE-AC02-98CH10886) for providing the facilities essential for 
the completion of this work.

\end{document}